\documentclass[aps,physrev,preprint,groupedaddress]{revtex4-2}
\usepackage{comment}
\usepackage{lineno}
\usepackage{svg}
\usepackage{graphicx} 

\begin{document}

%\preprint{}

\title{Modeling the Optical Properties of Biological Structures using Symbolic Regression}



\author{Julian Sierra-Velez}
\author{Alexandre Vial}
\author{Demetrio Mac\'{\i}as}
\email[Corresponding author: ]{demetrio.macias@utt.fr}
\affiliation{Laboratory Light, nanomaterials \& nanotechnologies - L2n, University of Technology of
Troyes \& CNRS UMR 7076, 12 Rue Marie Curie, 10004 Troyes, France}

\author{Marina Inchaussandague}
\author{Diana Skigin}
\affiliation{Universidad de Buenos Aires, Facultad de Ciencias Exactas y Naturales, Departamento de Física, Buenos Aires, Argentina}
\affiliation{CONICET - Universidad de Buenos Aires, Instituto de Física de Buenos Aires (IFIBA), Buenos Aires, Argentina}

\date{\today}

\begin{abstract}
We present a Machine Learning approach based on Symbolic Regression to derive, from either numerically generated or experimentally measured spectral data, closed-form expressions that model the optical properties of biological materials. To evaluate the performance of our approach, we consider three case studies with the aim of retrieving the refractive index of the materials that constitute the biological structures considered. The results obtained show that, in addition to retrieving readable and dimensionally homogeneous dispersion models, the expressions found have a physical meaning and their algebraic form is similar to that of the models often used to characterize the dispersive behavior of transparent dielectrics in the visible region.
\end{abstract}

\maketitle

%----------------------------------------------------------------------------------------
%----------------------------------------------------------------------------------------

\section{Introduction} \label{Chapters/intro} 
%----------------------------------------------------------------------------------------
%----------------------------------------------------------------------------------------

The development of increasingly powerful computational tools, combined with recent advances in machine learning (ML), has paved the way for data-driven approaches to tackle complex physical problems. In particular, Symbolic Regression (SR) has emerged as a robust methodology for discovering analytical expressions that model a given dataset~\cite{koza1994,Billard2002}. Contrary to traditional regression techniques, which assume a predefined functional form, SR explores a broad space of potential mathematical expressions without previous information on their algebraic form. This flexibility has positioned SR as a promising tool for scientific discovery, with researchers expecting it to unveil unknown underlying principles and relationships from experimental measurements~\cite{Angelis2023,arechiga2021}. 

In recent years, researchers have used SR to retrieve mathematical models that provide insight into the mechanisms that govern complex systems in different scientific disciplines, offering both predictive power and readability~\cite{Minh22,Schmidt2010, cranmer2020}. Despite its potential, traditional SR approaches face several challenges. The vast search space of potential expressions can make the optimization process computationally expensive, and relatively small changes in the training data may lead to different expressions~\cite{Kammerer2024}. Moreover, finding the optimal symbolic expression belongs to a class of problems for which no known algorithm can solve all instances efficiently (i.e., in polynomial time), making the search increasingly demanding as the problem size grows, and suggesting that SR itself might be an NP-hard problem~\cite{virgolin2022,Udrescu2020}. Furthermore,  there is an open discussion about the lack of physical interpretability, as the technique may fail to respect underlying principles such as dimensional homogeneity or conservation laws~\cite{Minh22,virgolin2022,Makke2024}. Modern approaches such as AI-Feynman~\cite{Udrescu2020} and PhySO~\cite{Tenachi2023,tenachi2024} address some of these limitations through advancements in computational algorithms and the integration of domain knowledge. 

We have previously explored the potential of SR to model the optical properties of transparent and absorbing dielectrics~\cite{Li2022} and also of transparent biological materials such as keratin and chitin~\cite{SierraVelez2023}. The results reported in those references showed the ability of SR to recover closed-form models from synthetic reflectance data. However, due to the operating principles of the numerical library used (see~\cite{Li2022} for details), they also made evident the impossibility to take into account the dimensional homogeneity and the algebraic complexity of the expressions searched. We have recently used the PhySO numerical library~\cite{Tenachi2023} to model the spectral reflectance of photonic biological structures made of disordered materials~\cite{SierraVelez2024}. The results reported in this reference settled the basis for the use of SR to model the optical properties of biological photonic structures. To the best of our knowledge, the performance of SR has not been explored in this context before. This presents a formidable challenge due to the complexity of biological structures.

%We have previously demonstrated the potential of SR to model the optical properties of transparent and absorbing dielectrics~\cite{Li2022} and in the context of biological photonic structures, where their optical response is inherently governed by complex physical principles~\cite{SierraVelez2024,SierraVelez2023}. These studies illustrated SR’s capability to retrieve closed-form models from synthetic and experimental reflectance data. However, its application to retrieving the optical properties of biological materials remains an open challenge. To our best knowledge, no prior research has explored the use of SR in this context. 

The structure of this paper is as follows. We introduce the biological structures to be studied and define their general geometric parameters in Section~\ref{Chapters/cases}. In Section~\ref{Chapters/methods} we describe the method used to compute the optical response of the structures and we detail our SR approach. In Section~\ref{Chapters/Results} we present and discuss the results of three cases treated in this work, using both synthetic and experimental data. At last, in Section~\ref{Chapters/conclu} we present our concluding remarks. 
%----------------------------------------------------------------------------------------
%----------------------------------------------------------------------------------------

\section{Statement of the problem} \label{Chapters/cases} 
%----------------------------------------------------------------------------------------
%----------------------------------------------------------------------------------------

Multilayer structures are responsible for the coloration of many biological photonic structures, where they generate vivid structural colors through light interference~\cite{Kinoshita2008,Sun2013}. Several animal species, including beetles, butterflies, and birds, feature nanostructures that resemble multilayer architectures that generate iridescence and dynamic color changes. Understanding the optical properties of these natural structures is a key step for bioinspired photonics, where these biological systems have already inspired applications in mechanics and optics~\cite{Tadepalli2017,Askarinejad2015}. 

\begin{figure}[ht!]
\centering
\includegraphics[width=0.6\linewidth]{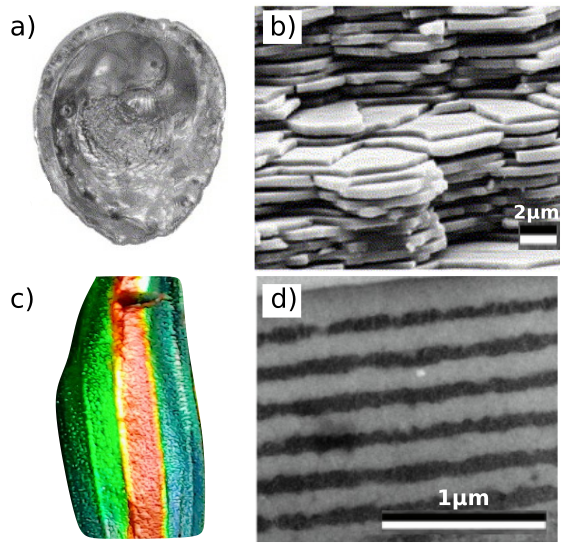}
\caption{Biological structures studied in this work. a) Image of the inner side of a shell composed of nacre. b) SEM image of a fractured cross-section of nacre, revealing its layered structure (images (a),(b) are adapted from Ref.~\cite{Kalpana2006}). c) Photo of our \textit{Jewel Beetle's} elytron sample. d) TEM image of the elytron's multilayer structure (adapted from Ref.~\cite{Yoshioka2011}).}
\label{fig:cases}
\end{figure}

In this work, we study two biological structures that exhibit interesting optical responses. The first, shown in Figs.~\ref{fig:cases}-a,b), is the nacre, a biomineralized composite material found in the inner layer of mollusk shells, that consists of a highly ordered arrangement of aragonite $\text{CaCO}_3$ platelets separated by organic layers. This alternating structure provides great mechanical strength and produces optical interference effects that give the shell a characteristic iridescent coloration~\cite{MADHAV2023101168}. The variations of the layers' thicknesses determine the observed hues, resulting in a color spectrum ranging from green to purple. This unique interplay of light and structure makes nacre an ideal candidate for biomimetic photonic system studies~\cite{Lertvachirapaiboon2015,Lertvachirapaiboon2018,Fan2021,Pina2013}. 

The second structure, shown in Figs.~\ref{fig:cases}-c,d), is the \textit{Chrysochroa} Jewel Beetle, known for its distinctive metallic green coloration due to the multilayer structure of its elytron~\cite{Noyes2007,Yoshioka2011,Stavenga2011,Yoshioka2012}. This consists of alternating chitin-melanic layers, arranged periodically, that create an iridescence effect~\cite{Stavenga2011}. 

\begin{figure}[ht!]
\centering
\includegraphics[width=0.5\linewidth]{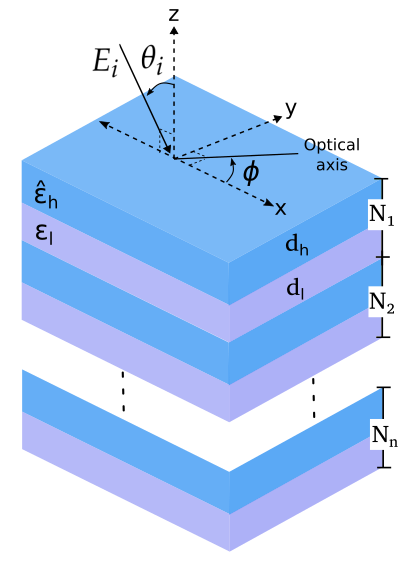}
\caption{Schematic representation of a general periodic $N_n$ bilayer stack of alternating $\epsilon_l$ and $\hat{\epsilon}_h$ permittivity layers, with thickness $d_l$ and $d_h$, respectively.}
\label{fig:multilayer}
\end{figure}

To model these biological structures, we consider the general periodic multilayer geometry depicted in Fig.~\ref{fig:multilayer}. The superstrate and substrate media are assumed to be air. The structure consists of a periodic $N_n$ bilayer stack of alternating $\epsilon_l$ and $\hat{\epsilon}_h$ permittivity layers, with thickness $d_l$ and $d_h$, respectively. We consider the material characterized by $\epsilon_l$ as isotropic, therefore, its optical properties are well described by the refractive index $n_l=\sqrt{\epsilon_l}$. Instead, the other material is characterized by its permittivity tensor $\hat{\epsilon}_h$, which enables us to consider anisotropic media.

For lossless anisotropic materials, $\hat{\epsilon}_h$ is a real, positive definite, symmetric tensor. Therefore, it can always be reduced to a diagonal form in an orthogonal coordinate system formed by its eigenvectors. In this work we consider uniaxial media. In such a case, two of the three diagonal elements (eigenvalues) are equal, and the permittivity tensor takes the form
\begin{equation}
\hat{\epsilon}_h = 
\left(
\begin{array}{ccc}
\epsilon_o & 0 & 0 \\
0 & \epsilon_e & 0 \\
0 & 0 & \epsilon_o
\end{array}\right),
\label{eq:epsilon_tensor}
\end{equation}
with $n_o=\sqrt{\epsilon_o}$ and $n_e=\sqrt{\epsilon_e}$ being the ordinary and extraordinary refractive indices, respectively. The optical axis is the unit eigenvector corresponding to the nonrepeated eigenvalue $\epsilon_e$.  For the purpose of keeping the complexity of the problem at a manageable level, we will assume that the optical axis is located in the $xy$ plane, forming an angle $\phi$ with the $x$-axis, as indicated in Fig.~\ref{fig:multilayer}. However, it is important to note that this simplified configuration was chosen for illustration purposes only and does not correspond to the general case of anisotropic biological materials that can be found in nature. 
%It lies on the $xy$ plane, forming an angle $\phi$ with the $x$ axis, as indicated in Fig.~\ref{fig:multilayer}. 
For an isotropic medium, $\epsilon_o=\epsilon_e=\epsilon_h$, and the material can be well described by its refractive index $n_h=\sqrt{\epsilon_h}$.  

As illustrated in Fig.~\ref{fig:multilayer}, the structure is illuminated in the $xz$ plane by an incident plane wave, denoted by $E_i$, at an angle $\theta_i$ relative to the surface's normal.

%----------------------------------------------------------------------------------------
%----------------------------------------------------------------------------------------

\section{Methodology} \label{Chapters/methods} 

%----------------------------------------------------------------------------------------
%----------------------------------------------------------------------------------------

In this section, we describe the methodology used to retrieve from reflectance spectra, the optical properties of the material characterized by $\hat{\epsilon}_h$ shown in the multilayer structure of Fig.~\ref{fig:multilayer}. First, we describe the method employed to compute the reflectance spectrum related to the structure considered. Then, we outline the SR approach used in this contribution.

\subsection{Optical Response of a Multilayer System}

We compute the optical response of multilayer systems using the Berreman $4\times4$ matrix method coupled with the S-matrix formalism. This approach ensures numerical stability and accounts for both isotropic and anisotropic materials~\cite{Berreman1972,stallinga1999,yeh1990,Langevin2024}. 

We consider the stratified structure presented in Fig.~\ref{fig:multilayer}, where the multilayer stack is invariant along the $x$ and $y$ directions, and the $z$-axis is the stacking direction. The incident plane is defined as the $xz$-plane. Assuming an implicit harmonic time dependence of the form $\exp(-i\omega t)$, Maxwell’s curl equations take the form
\begin{equation}
    \nabla \times \mathbf{E} = i\omega\mu_0 \mathbf{H}, \qquad 
    \nabla \times \mathbf{H} = -i\omega \hat{\epsilon} \mathbf{E},
\end{equation}
where $\hat{\epsilon}$ is the relative permittivity tensor as defined in Eq.~\ref{eq:epsilon_tensor}.

To solve these equations in a stratified medium, we assume a plane-wave solution with translational invariance in the $x$ direction. Specifically, we impose an $x$-dependence of the form $\exp(i k_x x)$ for all field components, where $k_x = k_0 \sin\theta_i$ is the conserved in-plane component of the wavevector, with $k_0=2 \pi / \lambda$. Since the system is also invariant along the $y$ direction, we set $k_y = 0$, and the problem reduces to solving for the field evolution along $z$ only.

Substituting this ansatz into Maxwell’s equations leads to a system of coupled first-order ordinary differential equations in $z$. These equations govern the behavior of the tangential field components, which are grouped into the four-component state vector $\mathbf{\Psi}(z)$ defined as
\begin{equation}
    \mathbf{\Psi}(z) =
    \left(
    \begin{array}{c}
    E_x(z) \\
    H_y(z) \\
    E_y(z) \\
    -H_x(z)
    \end{array}
    \right),
\end{equation}
representing a reduced-field formulation of Maxwell’s equations.

The evolution of $\mathbf{\Psi}(z)$ in each layer is governed by the Berreman matrix $\mathbf{M}$:
\begin{equation}
    \frac{d}{dz} \mathbf{\Psi}(z) = i \mathbf{M} \, \mathbf{\Psi}(z),
    \label{eq:M_matrix}
\end{equation}
where $\mathbf{M}$ is a $4\times4$ matrix that depends on $k_x$, $\omega$, and the dielectric tensor $\hat{\epsilon}$. The eigenvalues $\gamma_j$ of $\mathbf{M}$ correspond to propagation constants along $z$, and their associated eigenvectors $\mathbf{v}_j$ describe the mode profiles within the layer.

For isotropic media, $\mathbf{M}$ is block-diagonal and the TE and TM modes are decoupled. In contrast, for anisotropic media, off-diagonal components of the permittivity tensor introduce cross-coupling between electric field components, resulting in polarization conversion and mixing of TE and TM modes.

The general solution of Eq.~\ref{eq:M_matrix} in the $n$-th homogeneous layer is a linear combination of the four eigenmodes
\begin{equation}
    \mathbf{\Psi}_n(z) = \sum_{j=1}^{4} A_{n,j} \,\mathbf{v}_{n,j} \,e^{i \gamma_{n,j} z},
\end{equation}
where the coefficients \( A_{n,j} \) are the amplitudes of the eigenmodes. We define the total amplitude vector as \( \mathbf{A}_n = (A_{n,1}, A_{n,2}, A_{n,3}, A_{n,4})^T \), where $j=1,2$ correspond to the forward- and $j=3,4$ to the backward-propagating modes, respectively.

At each interface, the continuity of the tangential components of \( \mathbf{E} \) and \( \mathbf{H} \) provides the boundary conditions needed to relate the modal amplitudes across adjacent layers \( n-1 \) and \( n \). These relations are handled using the scattering matrix (S-matrix) formalism. To properly define the S-matrix, it is necessary to reorder the amplitudes according to physical propagation direction. This involves regrouping the modes into incoming and outgoing waves at each interface, based on the sign of the real part of their \( z \)-component of the Poynting vector. For each pair of adjacent layers, the partial matrix \( \mathbf{S}_{(n-1),n} \) relates the incoming \( \mathbf{A}_{in} = (A_{(n-1),1}, A_{(n-1),2}, A_{n,3}, A_{n,4})^T \) and outgoing \( \mathbf{A}_{out} = (A_{n,1}, A_{n,2}, A_{(n-1),3}, A_{(n-1),4})^T \) modal amplitudes across the interface as

\begin{equation}
    \mathbf{A}_{\text{in}} = \mathbf{S}_{(n-1),n} \, \mathbf{A}_{\text{out}}.
    \label{eq:s_matrix}
\end{equation}

To obtain the full optical response of the multilayer system, these partial matrices are combined recursively. The global system S-matrix is therefore constructed iteratively using the stable cascading rule for scattering matrices~\cite{Li1996}
\begin{equation}
    \label{eq:s_total}
    \mathbf{S}_{\text{total}} = \mathbf{S}_{0,1} \star \mathbf{S}_{1,2} \star \dots \star \mathbf{S}_{(N-1),N},
\end{equation}

where $\star$ denotes the cascading operation, which prevents numerical instabilities and properly accounts for multiple internal reflections and polarization mixing. From $\mathbf{S}_{\text{total}}$, one extracts the complex reflection coefficients.

%----------------------------------------------------------------------------------------
%----------------------------------------------------------------------------------------

\subsection{Symbolic Regression Approach and Implementations} \label{Chapters/SR} 

\begin{figure}[!ht]
\centering
\includegraphics[width=0.7\linewidth]{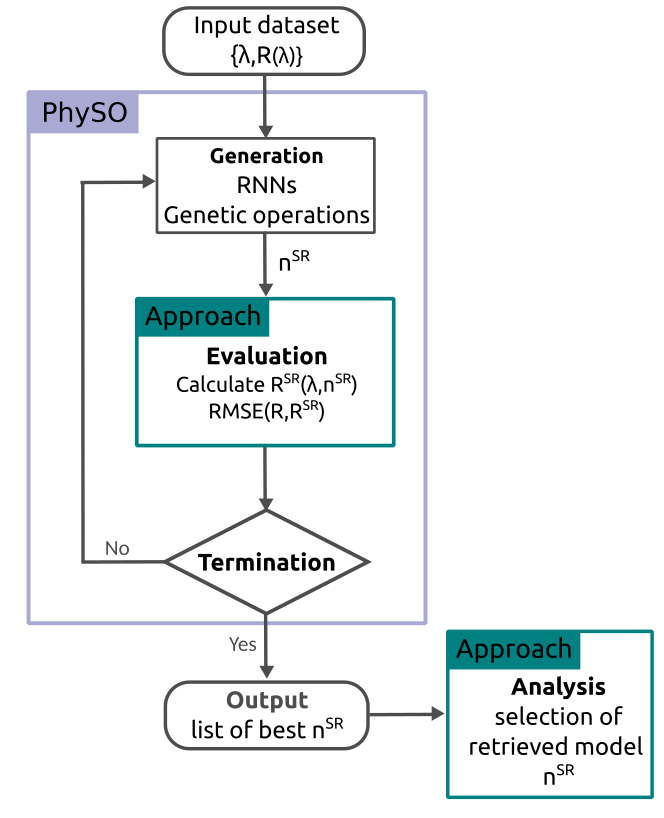}
\caption{Schematic representation of the SR framework incorporating our SR approach. The input dataset contains the incident wavelengths $\lambda$ and their corresponding reflectance values $R(\lambda)$. Candidate symbolic models $n^{SR}(\lambda)$ are generated using recurrent neural networks (RNNs) and genetic operations. These expressions are evaluated by computing the predicted reflectance $R^{SR}(\lambda, n^{SR})$ and comparing it to the target data via the RMSE metric. The process iterates until a termination criterion is met, yielding a list of expressions. A final analysis selects the most suitable retrieved model $n^{SR}$.}
\label{fig:physo_scheme}
\end{figure}

 In Fig.~\ref{fig:physo_scheme} we schematically present the general SR framework used in this contribution, to retrieve the closed-form expressions that characterize the dispersive behavior of the materials considered. 

SR is a general technique that, in recent years, has not only taken advantage of significant progress in the field of Machine Learning but has also been successfully employed across various scientific disciplines.
The flux diagram in Fig.~\ref{fig:physo_scheme} outlines the operating principles of the open-source library \textit{Physical Symbolic Optimization} (PhySO) used in this work. PhySO integrates dimensional analysis into the SR process and couples it with its reinforcement learning-based optimization. These two features make PhySO a robust and reliable state-of-the-art tool able to provide readable and dimensionally homogeneous models~\cite{Tenachi2023,tenachi2024}. For a deeper understanding, we refer the interested reader to the just-cited references. In the forthcoming paragraphs, we briefly outline the main stages that constitute PhySO's regression process.

\textbf{Define input dataset}: The process begins by defining the input dataset, which consists of the problem's variables (e.g., the wavelength $\lambda$) and the corresponding target values (e.g. the reflectance spectrum $R$). A critical feature of PhySO is its integration of dimensional analysis, requiring explicit specification of the physical dimensions of variables and target values. This ensures that all generated expressions remain dimensionally consistent. In this work, the units of the wavelength $\lambda$ are microns, and the target values correspond to the reflectance spectra $R(\lambda)$, which is dimensionless. In the studied cases involving nacre the input spectra are numerically generated, whereas in the Jewell Beetle case, the spectrum is experimentally measured.

\textbf{Specify Allowed Operations and Constants}: The next step involves defining the building blocks that PhySO is allowed to use for generating the predicted models $n^{SR}$. This is a set of mathematical operations, which could include arithmetic operations (e.g., addition, subtraction, multiplication, division), exponentials, logarithms, or trigonometric functions. Additionally, PhySO requires constants with specified dimensions to be included in the search space. The choice of the allowed building blocks is entirely dependent on the problem and the nature of the search.  Since the materials considered in this work are non-absorbing dielectrics, the four basic operations $\{+,-,*,/\}$ are sufficient to achieve our objectives. This choice was confirmed by extensive numerical experiments, not included here due to space limitations. These experiments showed that the SR could discriminate unnecessary operations and tended to retain only the basic ones in the final expressions.  Regarding the allowed constants, our rule of thumb is to use two constants for each variable with a matching dimension. In this work, we defined the set of allowed constants as $\{A(\mu\text{m}), B(\mu\text{m}), C, D\}$, where the first two match the units of $\lambda$, and the other two are dimensionless as the response. Note that there is an optimization process to find the numerical value of each one, therefore, each allowed constant adds considerable computing time.

\textbf{Expression Generation}: Using a combination of deep reinforcement learning and Recurrent Neural Networks (RNNs), PhySO generates symbolic expressions $n^{SR}$ iteratively. Starting with simple combinations of the allowed operations, the algorithm builds increasingly complex expressions. Dimensional homogeneity is ensured at every step by taking into account the defined units of the constants and variables in each operation. In this work, the predicted models correspond to the optical properties of the material characterized by $\hat{\epsilon}_h^{SR}$.

\textbf{Evaluation}: This step is the core of our approach, as it establishes the link between the SR process and the underlying physics of the problem. Given the predicted model $n^{SR}$ generated in the previous step, we use the Berreman 4×4 method to compute the reflectance spectrum $R^{SR}(\lambda,n^{SR})$. To evaluate its accuracy, we compare it with the target values $R$ using the Root Mean Square Error (RMSE) fitness metric
\begin{equation}
\text{RMSE}_R = \sum_{j=1}^{M} \sqrt{\frac{1}{N}   \sum_{i=1}^{N} \left( R_j(\lambda_i) - R^{SR}_j(\lambda_i,n^{SR}_j) \right)^2},
\label{eq:RMSE}
\end{equation}
where $M$ is the number of target reflectance spectra, $N$ is the total number of data points in the reflectance spectrum, and $i$ indexes each point $\lambda_i$ in the input variable domain. 

\textbf{Termination}: The evaluation step gives a list of predicted models $n^{SR}$, which are dimensionally homogeneous and readable models that describe the intrinsic property sought. If the best $n^{SR}$ does not meet a precision RMSE threshold, the process continues through iterative refinement in the Generation step, using genetic operations and optimization techniques. This cycle is called an epoch. This iterative process is repeated until either the precision criterion is met or a predefined limit (a total of 15 epochs in this work) is reached. 

\textbf{Analysis:} Once the SR process is complete, the resulting list of candidate expressions must be analyzed to identify the most appropriate model. A common strategy is to select the expression with the lowest error metric. However, we have noticed that a more systematic and reliable approach involves repeating the SR procedure multiple times under different initial conditions. This allows us to identify expressions that are consistently retrieved across runs. To implement this, we perform several independent realizations of the SR process, each using a different random seed for initialization. During the evaluation step, the accuracy metric $\text{RMSE}_R$ (Eq.~\ref{eq:RMSE}) is used to quantify how well the predicted spectra reproduce the target spectra. While this metric provides a measure of accuracy, it does not reflect the stability or reliability of the retrieved expressions. Therefore, we define two additional metrics to quantify the consistency of the models within all the results. The first metric is named Presence, and it corresponds to the number of realizations on which the model was retrieved. The second metric is named Occurrence, and it measures the relative frequency of the model compared to all acceptable retrieved expressions. These two metrics are used exclusively during the analysis phase and are not involved in the evaluation step. Together, they provide a robust framework for selecting reliable and repeatable models from the SR process.

%----------------------------------------------------------------------------------------
%----------------------------------------------------------------------------------------

\section{Results}\label{Chapters/Results}

%----------------------------------------------------------------------------------------
%----------------------------------------------------------------------------------------

This section presents the results obtained from applying the proposed SR approach to the biological structures shown in Fig.~\ref{fig:cases}. The results are divided into three study cases. In the first, we study the nacre multilayer structure assuming the aragonite layer to be an isotropic material. In the second case, we consider the same nacre structure but treat the aragonite layer as an anisotropic uniaxial material. Both cases are explored using numerically generated reflectance spectra.

The third case evaluates the performance of the SR approach when it is used to retrieve, from experimentally measured reflectance spectra, the dispersion model of one of the materials that constitutes the multilayer structure of  \textit{Chrysochroa} Jewel Beetle's elytron.

All over this section, we use the Berreman 4x4 method to compute numerically the predicted reflectance $R^{SR}(\lambda,n^{SR})$ of the structure studied. To this end, we use the open-source Python library \texttt{PyMoosh}~\cite{Langevin2024}. Furthermore, we only consider as acceptable candidate solutions those expressions with an associated metric $\text{RMSE}_R<0.05$. 

Extensive numerical experiments, not shown here due to space limitations, showed that increasing the number of input reflectance spectra for different angles of incidence did not necessarily enhance the robustness of the retrieved models. Also, we investigated the effect of varying the polarization state of the incident wave on the regression procedure. We observed that the results were consistent regardless of the polarization state of the input data considered, meaning that using one polarization state is sufficient to find a reliable solution. Based on these findings, we chose to use as the target data for each of the case studies previously described, only a single reflectance spectrum of the form
\begin{equation}
    R(\lambda) = \frac{1}{2}(R_{ss}(\lambda)+R_{pp}(\lambda)),
\end{equation}
where $R_{ss}$ and $R_{pp}$ are the total reflectance copolarized components for the s- and p-polarized incident fields, respectively.    

%----------------------------------------------------------------------------------------
%----------------------------------------------------------------------------------------
\subsection{Case: Nacre Multilayer with Isotropic Aragonite}\label{Chapters/caseA} 

%----------------------------------------------------------------------------------------
%----------------------------------------------------------------------------------------

In this first case study we model the nacre as the multilayer structure depicted in Fig.~\ref{fig:multilayer}. It consists of alternating layers of aragonite ($\text{CaCO}_3$), characterized by the refractive index $n_h$, and an organic material with $n_l = 1.55$. The thicknesses of the aragonite and organic layers are $d_h = 400~\text{nm}$ and $d_l = 50~\text{nm}$, respectively. In addition, the total number of periods is set to $N_n = 100$. The structure is illuminated by a plane wave at an incident angle of $\theta_i = 45^\circ$. To assess the performance of the SR-approach, we consider as a benchmark the refractive index of aragonite $n_h$ given by the Sellmeier equation~\cite{GHOSH1999}
\begin{equation}
n_h^2 (\lambda)= 1.734 + \frac{0.965 {\lambda^2}}{{\lambda^2} - 0.019 } + \frac{1.828 {\lambda^2}}{{\lambda^2} - 120},
\label{eq:aragonite}
\end{equation}
where $\lambda$ is the wavelength in micrometers.

The reflectance spectra that served as the target for the SR process were numerically generated over the visible wavelength range $ \lambda \in [0.375, 0.800]~\mu \text{m}$. These data were smoothed using a Gaussian filter with a standard deviation $\sigma = 2$ to minimize computational noise, without altering the overall spectral trends.

We performed 35 realizations, each corresponding to a different initialization seed, and the computing time per realization was about 1.9h. This expanded dataset provided a statistically significant set of predicted models to analyze the consistency of the retrieved expressions. The best retrieved models for the refractive index $ n_h^{SR} $ are summarized in Table~\ref{tab:iso_n}.

\renewcommand{\arraystretch}{1.5}
\begin{table}[!ht]
    \begin{ruledtabular}
    \caption{Best symbolic models retrieved for the isotropic refractive index of nacre  $ n_h^{SR} $. Each expression is reported with its Presence (P.) and Occurrence (O.) metrics across realizations, and accuracy metrics $ \text{RMSE}_n $, relative to the benchmark model, and $ \text{RMSE}_R $, relative to the numerically generated input data.}
    \label{tab:iso_n}
    \begin{tabular}{ccccc}
        \hline
        \textbf{Model} & \textbf{P.} & \textbf{O.} & \textbf{RMSE$_n$} & \textbf{RMSE$_R$} \\
        \hline
        $ n_{11} = \displaystyle a_{11} + \frac{b_{11}}{\lambda} $           & 35 & 35\% & 0.00126 & 0.03545 \\
        $ n_{12} = \displaystyle a_{12} + \frac{b_{12}}{\lambda + c_{12}} $  & 32 & 20\% & 0.00004 & 0.00151 \\
        $ n_{13} = \displaystyle a_{13} + \frac{b_{13}}{\lambda^2} $         & 31 & 20\% & 0.00018 & 0.01109 \\
        $ n_{14} = \displaystyle a_{14} + \frac{b_{14}}{\lambda + c_{14}} + \frac{d_{14}}{\lambda} $  & 13 & 5\% & 0.00013 & 0.00611 \\
        \hline
    \end{tabular}%
    \end{ruledtabular}
\end{table}

\begin{figure}[!ht]
    \centering
    \includegraphics[width=0.7\linewidth]{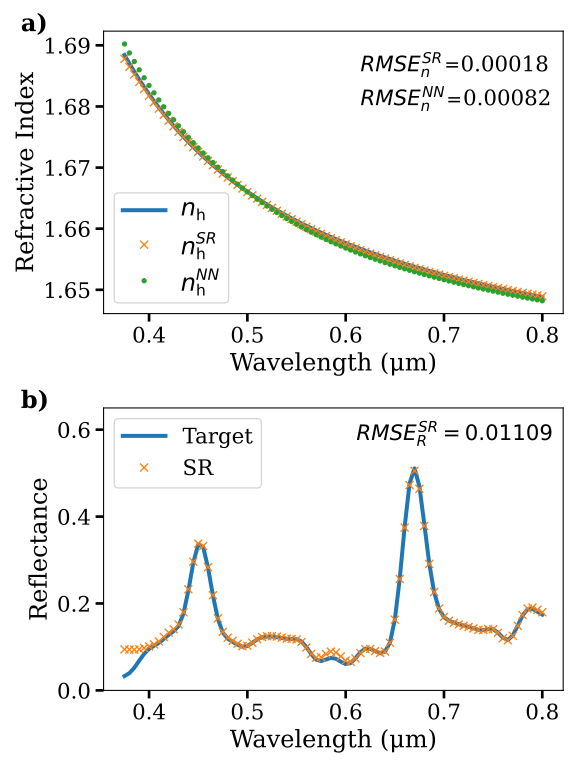}
    \caption{ a) Benchmark refractive index $ n_h(\lambda) $ given by Eq.~\ref{eq:aragonite}, compared with the predictions from SR ($ n_h^{SR} $) and NN ($ n_h^{NN} $). b) Reflectance spectra at a fixed incident angle $\theta_i=45^\circ$ for the input data (Target), and the computed spectrum using the predicted refractive index from SR (SR). }
    \label{fig:scenario2_results}
\end{figure}

Although any of the models presented in  Table~\ref{tab:iso_n} could be a suitable solution to the problem, the third one deserves special attention. Despite its relatively high $ \text{RMSE}_R $ metric values, its algebraic form resembles that of the well-established Cauchy model for the refractive index. This result suggests that our SR-approach is able to find a physically interpretable closed-form expression, among other suitable solutions that could be interpreted as local optima in the defined search space. Furthermore, it can be formally shown that the Cauchy and Sellmeier models provide equivalent results for transparent dielectrics within the visible range. This fact is illustrated in Fig. ~\ref{fig:scenario2_results}a), where we compare the benchmark refractive index given by the Sellmeier equation with the refractive index predicted by Eq.~\ref{eq:n_caseA}. 

\begin{equation}
    \centering
    n_h^{SR} (\lambda) = n_{13} = a_{13} + \frac{b_{13}}{\lambda^2},
    \label{eq:n_caseA}
\end{equation}
with $a_{13}=1.638$, and $b_{13}=0.007~\mu\text{m}^2$. 

For the sake of completeness, we also compare the results predicted by Eq.~\ref{eq:n_caseA} with those obtained with a Deep Neural Network (NN). For this numerical experiment, the network maps full reflectance spectra to the corresponding refractive index spectra over the visible range. It consists of an input layer with $N_\lambda$ neurons, three hidden layers with 256, 128, and 64 neurons, respectively, and an output layer with $N_\lambda$ neurons, where $N_\lambda$ denotes the number of sampled wavelengths. To train the NN, we first generated a synthetic database comprised of 10000 samples, each corresponding to a pair of a possible refractive index value with its associated reflectance spectrum. The generation of the database required nearly 2h of computation. The network was then trained on this dataset, which took 1.14h. After training, the model was used to predict the refractive index $n_h^{\text{NN}}$. In contrast, the SR approach required only a single reflectance spectrum as training, avoiding the need for a precomputed database.

Although the SR process is computationally intensive, with a higher computation time compared with the NN, it has the crucial advantage of providing a closed-form expression for the refractive index. While the NN model is accurate, it produces only numerical values without revealing explicit relationships within the data. Furthermore, in practical applications where only experimental measurements of the reflectance are available, the SR approach remains useful, unlike the NN method, which would require a full database that may not be experimentally accessible.

In Fig.~\ref{fig:scenario2_results}-b) we compare the reflectance spectra related to the SR model and the target reflectance data. As observed, the response of the SR model has a high prediction accuracy with a $\text{RMSE}_R = 0.01109$ metric value. Despite the small differences present for wavelengths below $0.4 \;\mu$m, the overall optical response predicted by the SR agrees with the target spectrum.

To further understand the operating principles of the SR approach, we repeated the previous numerical experiment keeping the same conditions. However, this time we searched for the closed-form expression of the dielectric function $\epsilon$ instead of searching for that of the refractive index $n$. This numerical experiment took about 1.3h per realization.

\renewcommand{\arraystretch}{1.5}
\begin{table}[!ht]
    \begin{ruledtabular}
    \caption{Best symbolic models retrieved for the isotropic dielectric function $ \epsilon_h^{SR} $ of nacre. Each expression is reported with its Presence (P.) and Occurrence (O.) metrics across realizations, and accuracy metrics $ \text{RMSE}_\epsilon $, relative to the square of the benchmark model, and $ \text{RMSE}_R $, relative to the numerically generated input data.}
    \label{tab:iso_eps}
    \begin{tabular}{ccccc}
        \hline
        \textbf{Model}  & \textbf{P.} & \textbf{O.} & \textbf{RMSE$_\epsilon$} & \textbf{RMSE$_R$} \\
        \hline
        $ \epsilon_{21} = \displaystyle a_{21} + \frac{b_{21}}{\lambda + c_{21}} $ & 35 & 49\% & 0.00007 & 0.00052 \\
        $ \epsilon_{22} = \displaystyle a_{22} + \frac{b_{22}}{\lambda} $          & 35 & 27\% & 0.00550 & 0.02137 \\
        $ \epsilon_{23} = \displaystyle a_{23} $                                   & 35 & 13\% & 0.03558 & 0.04971 \\
        $ \epsilon_{24} = \displaystyle a_{24} + \frac{b_{24}}{\lambda^2} $        & 11 & 4\% & 0.00205 & 0.03255 \\
        \hline
    \end{tabular}
    \end{ruledtabular}
\end{table}

The best retrieved models for $ \epsilon_h^{SR}$ are summarized in Table~\ref{tab:iso_eps}.  Although not physically interpretable, because of its fairly high and low respective Occurrence and $\text{RMSE}_R$ values, one may consider the first expression 

\begin{equation}
    \epsilon_h^{SR}(\lambda) = \epsilon_{21} = a_{21} + \frac{b_{21}}{\lambda + c_{21}},
    \label{eq:eps_caseA}
\end{equation}
with $ a_{21} = 2.657 $, $ b_{21} = 0.038 ~\mu\text{m} $ and $ c_{21} = -0.178 ~\mu\text{m} $, as the optimal solution. However, despite its fairly low Occurrence and high $\text{RMSE}_R$ values, the fourth expression in Table~\ref{tab:iso_eps} resembles the Cauchy model.
 
To explain this result, let us square the Cauchy model given by Eq.~\ref{eq:n_caseA}. After some simple algebra, we obtain 
\begin{equation}
    \centering
    \epsilon_h^{SR}(\lambda) = (n_h^{SR} (\lambda))^2 = a_{13}^2 + \frac{2a_{13}b_{13}}{\lambda^2} + \frac{b_{13}^2}{\lambda^4}.
    \label{eq:n2_caseA}
\end{equation}

The fourth equation in Table~\ref{tab:iso_eps} is similar to Eq.~\ref{eq:n2_caseA}, except for the third missing term. Furthermore, when we write the coefficients $a_{24}=a_{13}^2=2.682$, and $b_{24}=2a_{13}b_{13}=0.023~\mu\text{m}^2$ it is straightforward to see that the numerical value of $b_{13}^2$ makes the third term negligible with respect to the first two terms. 

The results from this first case study are encouraging, as they demonstrate the ability of SR to capture intrinsic relationships in the data. Furthermore, the underlying physical relation $ \epsilon_h = n_h^2 $ is still preserved. This highlights the robustness and consistency of our SR approach. However, we also found that expressions retrieved by SR with a physical meaning do not necessarily represent the optimal solution that minimizes the chosen regression metric. This interesting finding will repeatedly appear throughout the numerical experiments conducted in this study.

%----------------------------------------------------------------------------------------
%----------------------------------------------------------------------------------------

\subsection{Case: Nacre Multilayer with Anisotropic Aragonite} \label{Chapters/caseB} 

%----------------------------------------------------------------------------------------
%----------------------------------------------------------------------------------------
To assess further the performance of our SR scheme, we repeated the numerical experiment described in the previous case study, keeping the same geometrical and illumination conditions, but this time considering the optical anisotropy of the aragonite layer. To this end, in a first stage we searched for the expression of the ordinary index $ n_o $  assuming the extraordinary index $ n_e $ fixed. Then, in a second stage we inverted the configuration and searched for the expression of the extraordinary index $ n_e $ while fixing $ n_o $.

The target expressions for the aragonite layer's ordinary and extraordinary dielectric functions, $ \epsilon_o = n_o^2 $ and $ \epsilon_e = n_e^2 $, were respectively given by Sellmeier equations (Eq.~\ref{eq:aragonite}) and  
\begin{equation}
    n_e^2(\lambda) = 1.359 + \frac{0.824 \lambda^2}{\lambda^2 - 0.011} + \frac{0.144 \lambda^2}{\lambda^2 - 120},
\label{eq:extra}
\end{equation}
where $\lambda$ is the wavelength in micrometers. Furthermore, we assumed that the optical axis was on the $xy$ interface plane and that it was rotated by $ \phi = 45^\circ $ with respect to the $x$-axis.
 
We used numerically generated spectra as input data. Also, to avoid a significant increase in computing time, instead of the $N_n = 100$ periods initially considered we fixed them to $N_n = 10$. As in the first case study, we performed 35 realizations corresponding to different random seeds and the total computing time for each realization was about 3.9h.

The top four models found, for the two configurations described at the beginning of this case study, are respectively summarized in Tables~\ref{tab:aniso_no} and~\ref{tab:aniso_ne}. Each closed-form expression represents a possible dispersion model for the ordinary (resp. extraordinary) index $ n_o^{SR} $ (resp. $ n_e^{SR} $) and it is accompanied by its related metrics. 

\renewcommand{\arraystretch}{1.5}
\begin{table}[!ht]
    \begin{ruledtabular}
    \caption{Best symbolic models retrieved for the ordinary component $ n_o^{SR} $ of the anisotropic refractive index of nacre. Each expression is reported with its Presence (P.) and Occurrence (O.) metrics across realizations, and accuracy metrics $ \text{RMSE}_n $, relative to the benchmark model, and $ \text{RMSE}_R $, relative to the numerically generated input data.}
    \label{tab:aniso_no}
    \begin{tabular}{ccccc}
        \hline
        \textbf{Model} & \textbf{P.} & \textbf{O.} & \textbf{RMSE$_n$} & \textbf{RMSE$_R$} \\
        \hline
        $ n_{31} = \displaystyle a_{31} + \frac{b_{31}}{\lambda + c_{31}} $  & 32 & 40\% & 0.00004 & 0.00007 \\
        $ n_{32} = \displaystyle a_{32} + \frac{b_{32}}{\lambda^2} $  & 31 & 21\% & 0.00018 & 0.00032 \\
        $ n_{33} = \displaystyle a_{33} + \frac{b_{33}}{\lambda} $  & 22 & 10\% & 0.00126 & 0.00237 \\
        $ n_{34} = \displaystyle a_{34} + \frac{b_{34}}{\lambda + c_{34}} + \frac{d_{34}}{\lambda} $  & 12 & 5\% & 0.00013 & 0.00024 \\
        \hline
    \end{tabular}%
    \end{ruledtabular}
\end{table}

\renewcommand{\arraystretch}{1.5}
\begin{table}[!ht]
    \begin{ruledtabular}
    \caption{Best symbolic models retrieved for the extraordinary component $ n_e^{SR} $ of the anisotropic refractive index of nacre. Each expression is reported with its Presence (P.) and Occurrence (O.) metrics across realizations, and accuracy metrics $ \text{RMSE}_n $, relative to the benchmark model, and $ \text{RMSE}_R $, relative to the numerically generated input data.}
    \label{tab:aniso_ne}
    \begin{tabular}{ccccc}
        \hline
        \textbf{Model} & \textbf{P.} & \textbf{O.} & \textbf{RMSE$_n$} & \textbf{RMSE$_R$} \\
        \hline
        $ n_{41} = \displaystyle a_{41} + \frac{b_{41}}{\lambda^2} $  & 34 & 20\% & 0.00075 & 0.00120 \\
        $ n_{42} = \displaystyle a_{42} + \frac{b_{42}}{\lambda} $  & 33 & 15\% & 0.00060 & 0.00086 \\
        $ n_{43} = \displaystyle a_{43} + \frac{b_{43}}{\lambda + c_{43}} $  & 30 & 14\% & 0.00052 & 0.00073 \\
        $ n_{44} = \displaystyle a_{44} $  & 35 & 12\% & 0.00496 & 0.00747 \\
        \hline
    \end{tabular}
    \end{ruledtabular}
\end{table}

A remarkable feature observed in Tables~\ref{tab:aniso_no} and~\ref{tab:aniso_ne} is that in both of them the expressions for the ordinary and the extraordinary refractive indices of nacre, present the same algebraic form as those presented in Table~\ref{tab:iso_n}, corresponding to the isotropic case. Certainly, despite these similarities, the respective metrics differ. However, this is a somehow expected situation that can be directly related to the random nature of the SR-approach's operating principles. 

Another interesting fact observed in Tables~\ref{tab:aniso_no} and~\ref{tab:aniso_ne} is the presence of a Cauchy model-like expression, which once more shows the equivalence of Sellmeier and Cauchy models for transparent dielectrics in the visible range. Also, it is noteworthy to mention that Cauchy model in Table~\ref{tab:aniso_no} has the same coefficients as Eq.~\ref{eq:n_caseA}, that is $a_{32}=a_{13}=1.638$, and $b_{32}=b_{13}=0.007~\mu\text{m}^2$. In what concerns Cauchy model in Table~\ref{tab:aniso_ne}, its coefficients are $a_{41}= 1.477$, and $b_{41}=0.003~\mu\text{m}^2$ and according to its related metrics it is the best solution to this problem.

\begin{figure}[!ht]
    \centering
    \includegraphics[width=0.7\linewidth]{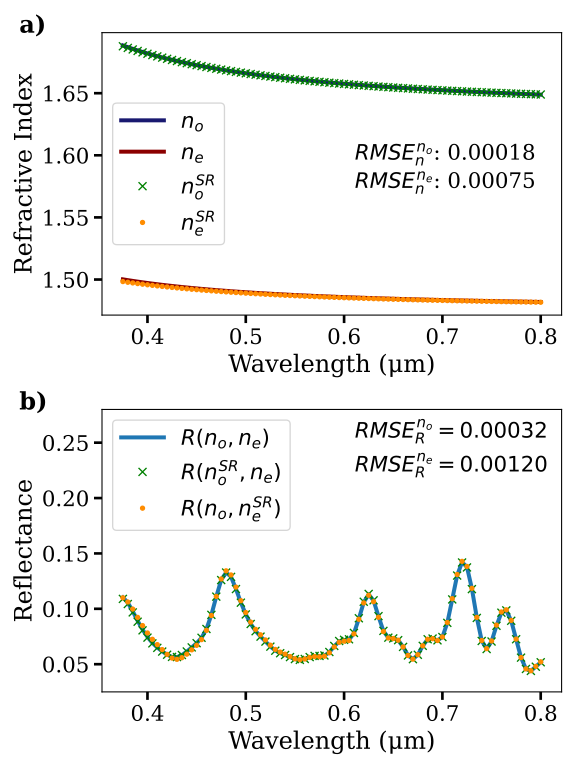}
    \caption{a) Benchmark ordinary $n_o$ and extraordinary $n_e$ refractive indexes, compared with the predicted $ n_o^{SR},n_e^{SR} $ models respectively. b) Reflectance spectra at $\theta_i=45^\circ$. The target data $R(n_o,n_e)$ is compared with the reflectance of the two numerical experiments' predictions,  $R(n_o^{SR},n_e)$ corresponding to fixing $n_e$ and SR predicting $n_o^{SR}$, and $R(n_o,n_e^{SR})$ by fixing $n_o$ and SR predicting $n_e^{SR}$. }
    \label{fig:ani_results}
\end{figure}

To better visualize these results, we compare in Fig.~\ref{fig:ani_results}-a) the refractive indices given by Eqs.~\ref{eq:aragonite} and~\ref{eq:extra} with their respective predictions provided by the SR-approach. For each case the curves are indistinguishable, a fact that is confirmed by the respective $\text{RMSE}_n$ metrics. 

Finally, in Fig.~\ref{fig:ani_results}-b) we present the target reflectance spectrum, depicted with a thick cyan solid line, together with the reflectance spectra generated with Cauchy-like models retrieved by the SR, for each of the two configurations considered. The three spectra are indistinguishable and, as in  Fig.~\ref{fig:ani_results}-a), this is confirmed by the respective $\text{RMSE}_R$ metrics.

%----------------------------------------------------------------------------------------
%----------------------------------------------------------------------------------------

\subsection{Case: \textit{Chrysochroa} Jewel Beetle Multilayer}\label{Chapters/caseC} 
%----------------------------------------------------------------------------------------
%----------------------------------------------------------------------------------------
Thus far, all of the presented case studies have used reflectance spectra as their input data. These spectra were generated using a well-established dispersion model, which allowed us to evaluate the performance of our SR approach objectively. However, there are often situations in which only experimentally measured input data is available and either no dispersion models exist or the existing ones are not suitable for the problem studied.

In this final case study, the goal is to evaluate the performance of our SR approach using reflectance spectra measured experimentally as input data. For this purpose, we consider the \textit{Chrysochroa} Jewel Beetle presented in Fig.~\ref{fig:cases}-c). 

\begin{figure}[!ht]
    \centering
    \includegraphics[width=0.65\linewidth]{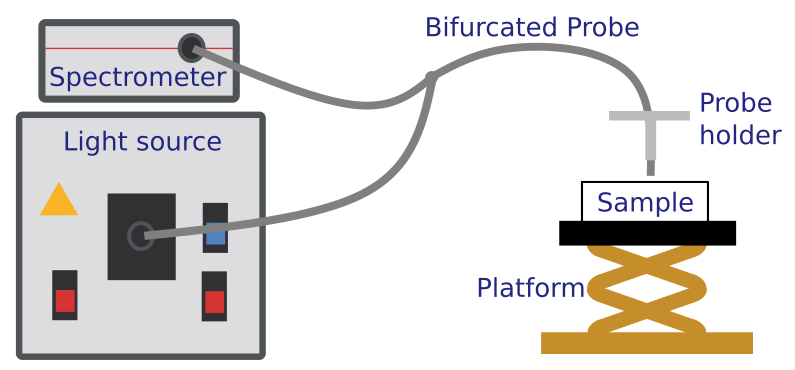}
    \caption{Optical setup used to measure the reflectance spectra of the Jewel beetle experimentally.}
    \label{fig:setup}
\end{figure}

The reflectance spectrum $ R $ of the Jewel beetle’s elytron was experimentally measured using the setup illustrated in Fig.~\ref{fig:setup}, where we employed a bifurcated optical fiber probe (Avantes FCR-7UV200-2-ME), coupled with an AvaSpec-2048 fiber optic spectrometer and an AvaLight-DH-S deuterium-halogen light source. The bifurcated probe enabled simultaneous illumination and collection of the reflected light from the sample, ensuring consistent measurement conditions. Measurements were performed under unpolarized light at normal incidence ($ \theta_i = 0^\circ $), with the probe oriented perpendicular to the sample surface. To ensure accurate alignment and reduce ambient light interference, the sample was mounted on a stable, adjustable platform, and the measurements were conducted in a controlled environment. The reflected signal captured by the spectrometer was recorded and subsequently used as input data for our retrieval approach.

Based on the optogeometric parameters reported in Ref.~\cite{Yoshioka2012}, the \textit{Chrysochroa} Jewel Beetle has been modeled as the multilayer structure shown in Fig.~\ref{fig:multilayer} with the layers' thicknesses set to $ d_l = 73~\text{nm} $ and $ d_h = 90~\text{nm} $, respectively. Also, the total number of periods was $ N_n = 5 $. Furthermore, the incident medium was air, and the semi-infinite substrate is also characterized by %$ n_l $.

\begin{equation}
    n_l(\lambda) = A + \frac{B}{\lambda^2},
    \label{eq:Cauchy}
\end{equation}
where $ A = 1.51 $ and $ B = 0.0153~\mu\text{m}^2 $.

We conducted a numerical experiment analogous to the previous cases, performing 35 realizations of the SR process. The total computation time for each realization was approximately 1.23h. 

The top retrieved models for the refractive index $ n_h^{SR} $ and their respective metrics are summarized in Table~\ref{tab:case_beetle}. Each expression is presented as an algebraic simplification of the output generated by PhySO. Also, since no benchmark model is available for $n_h$, the metric $\text{RMSE}_n $ is not included. 

\renewcommand{\arraystretch}{1.5}
\begin{table}[!ht]
    \begin{ruledtabular}
    \caption{Best symbolic models retrieved for $ n_h $. Each expression is reported with its Presence and Occurrence metrics across realizations, and the accuracy metric $ \text{RMSE}_R $ relative to the experimentally measured target data.}
    \label{tab:case_beetle}
    \begin{tabular}{cccc}
        \hline
        \textbf{Model} & \textbf{Presence} & \textbf{Occurrence} & \textbf{RMSE$_R$} \\
        \hline
        $ n_{51} = \displaystyle a_{51} + \frac{b_{51}}{\lambda + c_{51}} $         & 33 & 29\% & 0.02238 \\
        $ n_{52} = \displaystyle a_{52} $                                           & 35 & 23\% & 0.02467 \\
        $ n_{53} = \displaystyle a_{53} + b_{53} \lambda + \frac{c_{53}}{\lambda} $ & 31 & 18\% & 0.02395 \\
        $ n_{54} = \displaystyle a_{54} + \frac{b_{54}}{\lambda^2} $                & 2  & 0.6\% & 0.02686 \\
        \hline
    \end{tabular}
    \end{ruledtabular}
\end{table}

As could be expected, the $\text{RMSE}_R$ related to all the models in Table~\ref{tab:case_beetle}  are greater than any of the metrics associated with the expressions obtained in the previous cases.
This result does not necessarily imply a failure of our SR approach.  Rather, it can be explained by the fact that biological multilayer structures, such as the elytra of the Jewel beetle, often have non-ideal features, such as surface roughness, curvature, and variations in layer thickness. These features deviate from the perfectly flat, periodic geometry assumed in Fig.~\ref{fig:multilayer}. Consequently, rather than recovering an idealized dispersion relation, the SR approach searches for the best numerical fit to the measured optical response of the studied structure.

Remarkably, despite an algebraic form far from any well established dispersion model, the first expression in Table~\ref{tab:case_beetle} not only was retrieved in all the case studies presented in this work; but also, because of its related metrics, it provides the best approximation for the material under the assumed geometry. In the present case it is written as 
\begin{equation}
    n_h^{SR}(\lambda) = n_{51} = \displaystyle a_{51} + \frac{b_{51}}{\lambda + c_{51}},
    \label{eq:nh_caseB}
\end{equation}
with $ a_{51} = 2.483 $, $ b_{51} = 1.969 ~\mu\text{m} $, and $ c_{51} = -2.964 ~\mu\text{m} $.  

Another noteworthy result is the second model $ a_{52} = 1.68$ in Table~\ref{tab:case_beetle}, which corresponds to the same value reported in Ref.~\cite{Yoshioka2011} when $n_h$ is assumed constant. Furthermore, notwithstanding its "poor" metrics, other interesting result in Table~\ref{tab:case_beetle} is the Cauchy-like equation given by the fourth expression, whose coefficients are given by $ a_{54} = 1.56 $ and $ b_{54} = 0.036~\mu\text{m}^2 $, which correspond to the same values reported in reference~\cite{Yoshioka2011}. These results show that the SR can find physically interpretable expressions, without making any prior assumption concerning their algebraic form, using as input data experimentally measured spectral information. 

\begin{figure}[!ht]
    \centering
    \includegraphics[width=0.7\linewidth]{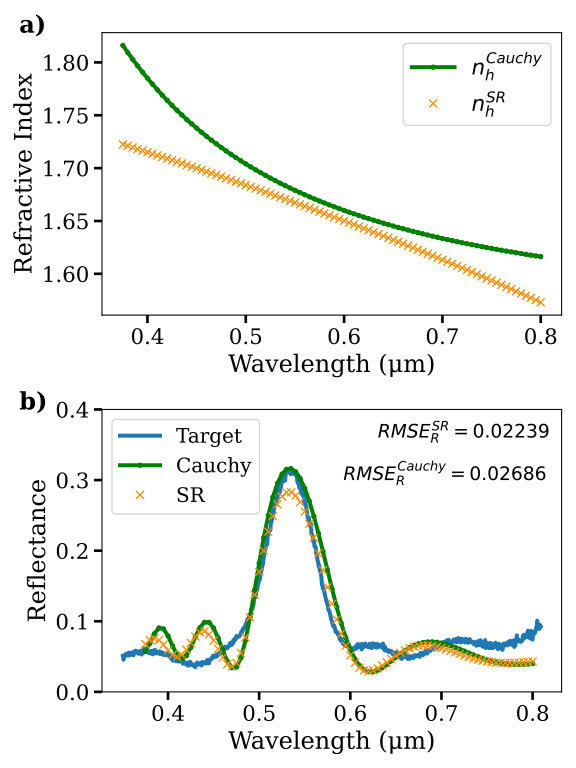}
    \caption{a) Comparison of the reported refractive index $ n_h^{Cauchy} $, and the SR prediction $ n_h^{SR} $. b) Reflectance spectra at normal incidence for experimentally measured data (Target), the SR prediction response $R(n_h^{SR})$, and the reported model response $R(n_h^{Cauchy})$.}
    \label{fig:caseB}
\end{figure}

To better visualize the performance of the retrieved model, we depict in Fig.~\ref{fig:caseB}-a) the reported refractive index $n_h^{Cauchy}$, with the model of $n_h^{SR}$ predicted by the SR through Eq.~\ref{eq:nh_caseB}. The effect of the obvious differences between the two models can be seen in Fig.~\ref{fig:caseB}-b), where we plot their related reflectance spectra.  It is clear that the two predicted models reproduce fairly well the central peak of the experimental spectrum. However, none of them is able to accurately reproduce the spectrum within the spectral range considered. Furthermore, despite its "lower" $\text{RMSE}_R$ metric, we observe that the amplitude of the peak predicted by Eq.~\ref{eq:nh_caseB} is slightly lower than that predicted by the Cauchy model.  
As stated previously, this apparent failure of our SR approach is more related to the simplified model used to represent the Jewell beetle's elytron, than to the SR operating principles themselves. Very likely, using a more sophisticated model or a numerical method such as the Finite Differences Time Domain or the Finite Elements Method could improve the accuracy of the predictions. Nevertheless, because of the SR's iterative nature, any of those methods would prohibitively increase the computing time.

Despite its simplicity, this last case study clearly illustrates the usefulness of SR when only experimental measurements are available. The results demonstrate that it is possible to obtain a closed-form expression that predicts the property under study within the considered spectral range, even with limited input data. It should be noted that this example is more complex because the model of the property being studied is obtained indirectly from the measured response.
	 	 	 	
On the other hand, this case study also reveals two important flaws of SR that should be considered when using it. First, while the final expression will be readable and dimensionally homogeneous, it may not have physical meaning because this characteristic does not imply the best fit of the chosen metric. Second, SR depends entirely on the geometric model used to represent the structure under study and the method used to calculate the response that fits the experimental data. In the presented case study, the multi-layered structure with flat interfaces is clearly a simplified representation of the beetle wing structure under consideration. Therefore, this lack of information inevitably influences the algebraic form of the final expression.

%----------------------------------------------------------------------------------------
%----------------------------------------------------------------------------------------

\section{Conclusions}\label{Chapters/conclu} 
%----------------------------------------------------------------------------------------
%----------------------------------------------------------------------------------------

The results of this study demonstrate the effectiveness of SR in modeling the optical properties of biological materials. 

It should be noted that the only information used by the SR in this work were the units of the considered variables and the far-field reflectance spectra, which were generated numerically or measured experimentally. The closed-form, dimensionally homogeneous expressions recovered by SR demonstrate the robustness of this tool. They also open the possibility of modeling the optical properties of more complex materials or structures without the need for a large database.

Unlike most references that use SR for a specific application, we evaluated the overall performance of our approach  by carrying out multiple realizations of it. In other words, we repeated the regression process starting from different initial states while keeping the illumination conditions and geometric parameters of the studied structures constant. As expected, SR does not provide a single dispersion model but rather a set of closed-form expressions modeling the dispersive behavior of the studied material.  However, our results showed that optimal solutions were often repeated throughout the realizations. Furthermore, the solutions included physically interpretable models, such as the Cauchy model, which has been shown to be valid for non-absorbing materials in the visible spectrum.

The results presented are encouraging and confirm that SR can be considered a transparent box that explicitly establishes the relationship between different data related to the studied physical phenomenon. In this sense, the potential applications of SR extend beyond the study of biological or biomimetic structures. However, much work is still needed, as many questions remain to be answered. Examples include handling complex functions, such as those of dispersive models of metallic or dielectric absorbing materials, and the multi-objective symbolic regression approach, which could address multi-physics problems.

%----------------------------------------------------------------------------------------
%----------------------------------------------------------------------------------------

%----------------------------------------------------------------------------------------
%----------------------------------------------------------------------------------------
\begin{acknowledgments}

This work was conducted within the framework of the Graduate School NANO-PHOT (École Universitaire de Recherche, PIA3, contract ANR-18-EURE-0013). 

M.I. and D.S. acknowledge partial support from Universidad de Buenos Aires (UBACyT 20020190100108BA) and CONICET (PIP 11220210100299CO).

We thank Marco A. Giraldo from the University of Antioquia, Colombia, for providing the sample of the \textit{Chrysochroa} Jewel Beetle elytron, and the setup for its reflectance measurement.

We thank Hendrik H{\"o}lscher from the Karlsruhe Institute of Technology (KIT), Germany, for fruitful discussions and reading of the manuscript. 
\end{acknowledgments} 

\textbf{Data availability} All data are available on reasonable request from the authors.

\textbf{Disclosures} The authors declare no conflicts of interest related to this article.

%----------------------------------------------------------------------------------------
%----------------------------------------------------------------------------------------

\bibliography{bibSR}

\end{document}